\documentclass[%
 aip,
 amsmath,amssymb,
 reprint,%
]{revtex4-1}

\usepackage{graphicx}
\usepackage{dcolumn}
\usepackage{bm}

\usepackage[T1]{fontenc}
\usepackage{mathptmx}
\usepackage{psfrag}

\begin{document}

\preprint{AIP/123-QED}

\title{\emph{Ab initio} calculations of structural stability, thermodynamic and elastic properties of Ni, Pd, Rh, and Ir at high pressures}

\author{N. A. Smirnov}
\email{nasmirnov@vniitf.ru}
\affiliation{Federal State Unitary Enterprise, Russian Federal Nuclear Center - Zababakhin All-Russian Research Institute of Technical Physics, 456770, Snezhinsk, Russia}

\date{\today}

\begin{abstract}
The paper presents results of a comprehensive study from first principles into the properties of Ni, Pd, Rh, and Ir crystals under pressure. We calculated elastic constants, phonon spectra, isotherms, Hugoniots, sound velocities, relative structural stability, and phase diagrams. It is shown that in nickel and palladium under high pressures ($>$0.14 TPa) and temperatures ($>$4 kK), the body-centered cubic structure is thermodynamically most stable instead of the face-centered cubic one. Calculated results suggest that nickel under Earth-core conditions ($P$$\sim$0.3 TPa, $T$$\sim$6 kK) have a bcc structure. No structural changes were found to occur in Rh and Ir under pressures to 1 TPa at least. The paper also provides estimations for the pressure and temperature at which the metals of interest begin to melt under shock compression.
\end{abstract}

\maketitle

\section{Introduction}

Persistent interest to the study of materials properties under high pressures and temperatures is determined, on the one hand, by progressive advances in the experimental methods used to study materials under these conditions and, on the other hand, by the discovery of interesting physical effects, for example, superconductivity at near-room temperatures. These studies also help judge the state of materials in the planetary interiors and gain a better understanding of processes in matter in the interest of inertial confinement fusion. Worthy of note is also the fact that the gradual growth of pressure (up to 1 TPa) accessible in experiments on static compression in diamond anvil cells both at room and at high temperatures \cite{A01,A02} raises the issue related to the appropriate pressure standard for strongly compressed matter. Some noble metals that were earlier used as the pressure standard have recently been found to exhibit polymorphism \cite{A03,A04,A05,A06,A07,A08} under high pressures and temperatures thus limiting their use for that purpose.

\emph{Ab initio} calculations into the behavior of Ni, Pd, Rh, and Ir under extreme conditions are relatively poor. The most thoroughly studied is nickel because its behavior under high $P$ and $T$ is of interest in research of planetary interiors. Its structural stability under pressure and zero temperature was studied in theoretical works \cite{A09,A10,A11}. The authors of papers \cite{A09,A10} performed calculations to pressures about 0.3 TPa and did not discover any structural changes. Much higher pressures were reached in calculations \cite{A11}, which showed that at $P$, above 6.3 TPa and $T$$=$0 K, face-centered cubic Ni must transform into a hexagonal close-packed structure. In turn, static experiments \cite{A12} did not reveal any structural transformations up to 0.37 TPa at room temperature.

Phonon spectra, thermodynamic and magnetic properties of nickel up to 0.1 TPa were calculated in paper \cite{A13}. The study showed that the magnetic fcc phase remained dynamically stable and its magnetic moment gradually reduced as compression grew. The evolution of the magnetic properties of Ni at higher pressures was investigated theoretically and experimentally in Refs.~\cite{A11,A14,A15}. Experiments \cite{A14,A15} suggest that the metal remains ferromagnetic at least to $P$$=$0.26 TPa. Calculations \cite{A11} estimate the pressure at which Ni completely loses its magnetic properties to be about 1 TPa.

Special attention is given to the melting curve of nickel. Data from early static experiments in laser-heated diamond anvil cells \cite{A16,A17} (laser-speckle method for detection of melting) disagree with \emph{ab initio} calculations \cite{A18,A19} and shock data \cite{A20}. The temperatures of melting at $P$ above 20 GPa reported in Refs.~\cite{A18,A19,A20} are markedly higher than in experiments \cite{A16,A17}. But new results of static measurements \cite{A21,A22} taken with an experimental technique (X-ray diffraction, XAS) different from that used in Refs.~\cite{A16,A17} show excellent agreement with data from \cite{A18,A19,A20}. That is, we can see that the optical method used in measurements \cite{A16,A17} does not give correct values for the melting temperature of nickel at high pressures.

The relative stability of Pd and Rh was studied from first principles in papers \cite{A23,A24}. These calculations show the fcc phase of Pd and Rh to be thermodynamically most stable at least to pressures about 1 TPa and $T$$=$0 K. Experimental studies of structural stability at room temperature below 80 GPa also show no phase transitions \cite{A25,A26,A27}. The melting curves of Pd and Rh were measured in papers \cite{A17,A28} under rather low pressures $<$ 30 GPa.

Iridium at high pressures (up to 0.26 TPa) and room temperature was studied in static experiments \cite{A29,A30,A31}. In experiment \cite{A29}, additional peaks were found in its diffraction pattern at relatively low $P$ which pointed, in the authors' view, to a transition into a new hexagonal structure. But later experiments did not confirm that \cite{A30,A31}. It seem so that no structural transitions occur in iridium at room temperature to pressures about 0.26 TPa. Polymorphism in Ir was not also seen in laser-heated diamond anvil cell experiments at $T$$\leq$3.1 kK and $P$$<$50 GPa \cite{A32}. Experimental data on the melting curve of Ir under pressure are very poor. The only estimation for its melting point at about 40 GPa is provided in experiments \cite{A32}.

The relative stability of Ir phases was theoretically studied in papers \cite{A33,A34}. Calculations \cite{A33} did not predict any structural changes to occur under pressures below $\sim$0.1 TPa, but the authors of article \cite{A34} found a thermodynamic stability region for the random-stacking hexagonal close-packed structure (rhcp) at high pressures and above-room temperatures. It is claimed that at $P$$=$0.2 TPa, the fcc$\rightarrow$rhcp transition must occur at $T$$\approx$1.6 kK. With the increasing pressure, the rhcp existence region on the phase diagram gradually extends to lower temperatures \cite{A34}. It is also predicted that the Hugoniot cross the fcc-rhcp phase boundary at $P$$\approx$0.16 TPa. The authors of article \cite{A34} demonstrate that the double hexagonal close-packed (dhcp) structure of Ir also has a lower energy than its fcc phase at high pressures and temperatures. In the recent paper \cite{A35} an equation of state (EOS) for Ir under pressures up to 0.54 TPa and temperatures up to 3 kK was constructed from quantum molecular dynamics (QMD) calculations but the structural stability of Ir was not considered.

This paper presents \emph{ab initio} calculations for the relative structural stability for four metals $-$ Ni, Pd, Rh, and Ir $-$ under high pressures and temperatures. The $PT$-diagrams of Ni, Pd, Rh, and Ir constructed from calculated results are provided. The existence of polymorphism in some of them (Ni, Pd) under pressures above hundred GPa is discussed. Thermodynamic and elastic properties of the four metals are calculated in a wide range of compressions and the dependencies of sound velocities on pressure under shock loading are determined. The paper provides estimations for the pressure and temperature at which the metals of interest begin to melt under shock compression.

\section{Calculation method}

In this work calculations were done with the well-known linear muffin-tin orbital method FP-LMTO \cite{A36} we have successfully used earlier \cite{A03,A08}. The method calculates the energy of a crystal as a function of specific volume $V$ and the degree of lattice deformation. Also, the FP-LMTO method can be used to calculate the phonon spectra of crystals at $T$$=$0 K with the help of linear response theory. With this method we determined the dependence of the energy of our metals versus compression for several types of lattices at zero and nonzero electron temperatures. The dependence of pressure on volume was determined through differentiation of an analytical expression that approximated \cite{A37} the calculated dependence of energy versus volume.

To determine how the heating of the electron subsystem contributes to the free energy of a crystal it is necessary to calculate electron gas entropy. It is calculated as \cite{A38}
\begin{equation}
S_e(T_e)=-k_B\int\limits_{-\infty}^{\infty}N(\varepsilon)[f_e\ln(f_e)+(1-f_e)\ln(1-f_e)]d\varepsilon, \label{eq001}
\end{equation}
Here $f_e$ is the Fermi-Dirac function, and $N$($\varepsilon$) is the electron density of states. With the known entropy and internal energy $E_e$ of electrons it is easy to obtain the free energy of the heated electron gas, $F_e$$=$$E_e$$-$$TS_e$.

Lattice vibrations are taken into account in a quasiharmonic approximation with the calculated phonon spectra. The phonon contributions to free and internal energies are written as \cite{A38}
\begin{equation}
F_{ph}=k_{B}TV\int\limits_{0}^{\infty}\ln\left[2\sinh\left(\frac{\hbar\omega}{2k_{B}T}\right)\right]g_{ph}(\omega,V)d\omega, \label{eq002}
\end{equation}
\begin{equation}
E_{ph}=\frac{1}{2}V\int\limits_{0}^{\infty}\hbar\omega\coth\left(\frac{\hbar\omega}{2k_{B}T}\right)g_{ph}(\omega,V)d\omega, \label{eq003}
\end{equation}
where $g_{ph}$ is the phonon density of states (PDOS), and $\omega$ is phonon frequency. The well-known Lindemann criterion was used to evaluate the position of the melting curve for the metals of interest. The procedure used to determine the curve is described in Refs.~\cite{A19,A39}. As shown in various calculations \cite{A03,A08,A19,A39,A40}, the Lindemann criterion performs rather well for different metals.

The valence electrons were 3$s$, 3$p$, 3$d$, 4$s$ for nickel, 4$s$, 4$p$, 4$d$ for palladium, 4$s$, 4$p$, 4$d$, 5$s$ for rhodium, and 5$s$, 5$p$, 4$f$, 5$d$, 6$s$ for iridium. FP-LMTO parameters were chosen so as to ensure high accuracy of \emph{ab initio} calculations (internal energy accurate to $\sim$0.1 mRy/atom). Integration over the Brillouin zone was done with the improved tetrahedron method \cite{A41}. Meshes in k-space measured 30$\times$30$\times$30 for the cubic structures and 30$\times$30$\times$18 for the hexagonal ones. For the determination of phonon frequencies, meshes over q-points were 10$\times$10$\times$10 for the cubic structures and 10$\times$10$\times$6 for the hexagonal ones. The cutoff energy for representing the basis functions as a set of plane waves in the interstitial region was scaled with respect to compression but was no lower than 900 eV. The basis set included MT-orbitals with moments to $l^b_{max}$$=$3. Charge density and potential expansions in terms of spherical harmonics were done to $l^w_{max}$$=$7. The $c/a$ parameter of hexagonal structures was always optimized. The other FP-LMTO parameters were chosen with an approach similar to that one described in article \cite{A03}.

\begin{figure}
\centering{
\includegraphics[width=8.0cm]{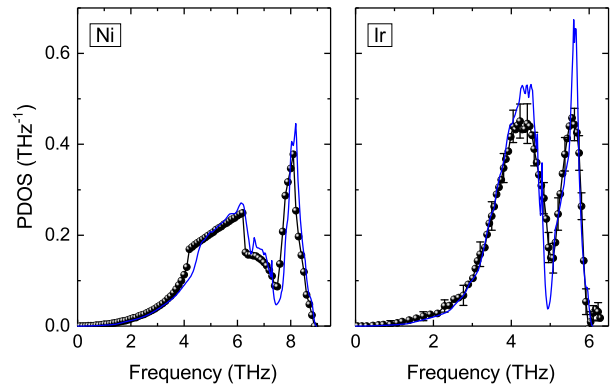}}
\caption{\label{fig1} Phonon densities of states for nickel and iridium in comparison with experimental data \cite{A45,A46}. The blue lines show our calculations at $T$$=$0 K and the black circles are data from room-temperature experiment.}
\end{figure}

The exchange-correlations (XC) functionals were chosen so as to attain the best description of ground state properties and phonon spectra for the metals.  The PBE functional \cite{A42} was taken for nickel and the PBEsol \cite{A43} was found best for the others. For all the elements, the equilibrium specific volume $V_0$ under ambient conditions was reproduced accurate to better than 1\% compared to experiment. The accuracy of our phonon spectrum calculations is demonstrated in Fig.~\ref{fig1} showing the spectra calculated in this work for nickel and iridium in comparison with experiment. They are seen to agree well.  A similar comparison for palladium was earlier demonstrated in our paper \cite{A44}. Experimental phonon spectrum data for rhodium are absent.

Elastic constants were calculated with an approach \cite{A47} based on the calculation of specific energy as a function of the degree of single-crystal deformation. The elastic constants obtained for single crystals were used to calculate longitudinal and bulk sound velocities versus applied compression for polycrystals by Voigt-Reuss-Hill averaging \cite{A48}. More details about these calculations can be found in paper \cite{A49}.

\section{Calculated results}

\begin{figure}
\centering{
\includegraphics[width=8.0cm]{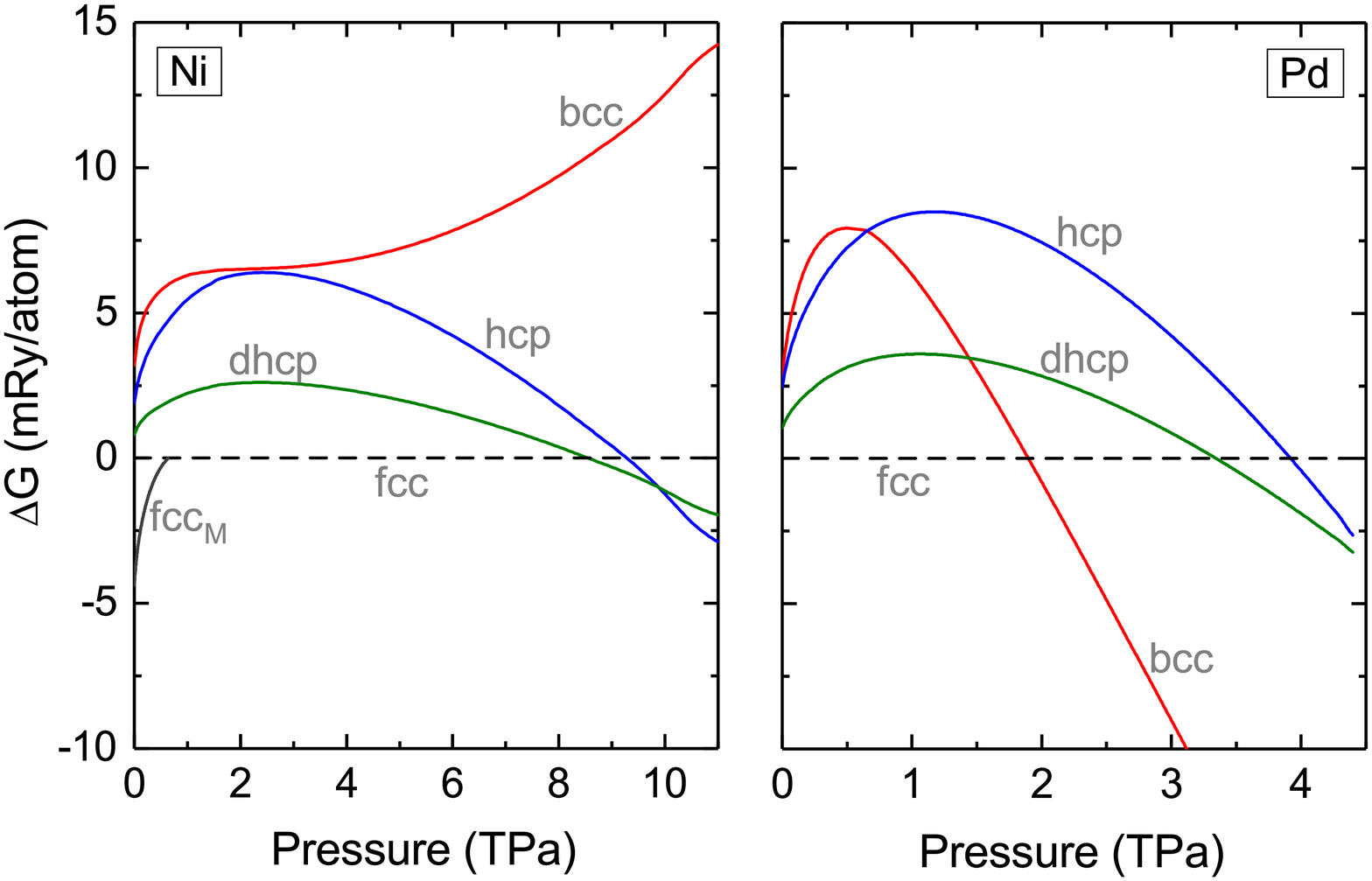}}
\caption{\label{fig2} Gibbs potential differences versus pressure between Ni and Pd structures at $T$$=$0 K.}
\end{figure}

\begin{figure}
\centering{
\includegraphics[width=8.0cm]{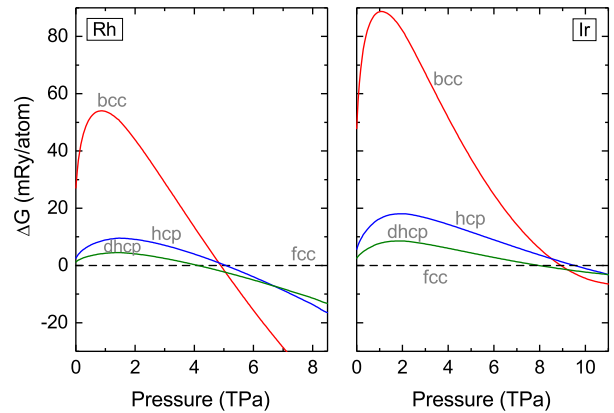}}
\caption{\label{fig3} Gibbs potential differences versus pressure between Rh and Ir structures at $T$$=$0 K.}
\end{figure}

Let's consider the relative stability of various Ni, Pd, Rh, and Ir crystal phases at zero temperature. Figures~\ref{fig2} and~\ref{fig3} show Gibbs potential differences versus pressure between different structures of the metals relative to the fcc potential. It is seen that with the increasing compression the magnetic fcc structure of nickel (Fig.~\ref{fig2}) which is most stable at low $P$ transforms into the non-magnetic fcc phase at $P$$>$0.6 TPa. Further compression results in its transition to dhcp ($\sim$8.5 TPa) and then to hcp ($\sim$9.9 TPa) structures. We note here that the bcc phase at $T$$=$0 K is energetically least preferable compared to the other close-packed phases but at $P$$\lesssim$2 TPa the difference between energies is relatively small ($\lesssim$6 mRy/atom). As for palladium, its compression to a few TPa shows only one structural transition to the bcc phase. Its behavior is very similar to that of platinum we have earlier studied in paper \cite{A08}. The pressure of the fcc$\rightarrow$bcc transition in Pd equals 1.9 TPa which is close to the value of about 2 TPa for the similar transition in platinum.

Rhodium and iridium at $T$$=$0 K (Fig.~\ref{fig3}) undergo the same sequence of structural transitions under pressure: first fcc$\rightarrow$dhcp and then dhcp$\rightarrow$bcc. The pressures of these transitions are about 4.1 and 5 TPa for Rh, and about 8 and 9.2 TPa for Ir, respectively. Note that in three (Pd, Rh, Ir) of the four metals at high compression and $T$$=$0 K, the bcc structure becomes energetically more preferable than the other close-packed phases. But it will be shown below that the bcc phase of nickel can also be present on its phase diagram even at relatively low pressures.

\begin{figure}
\centering{
\includegraphics[width=8.0cm]{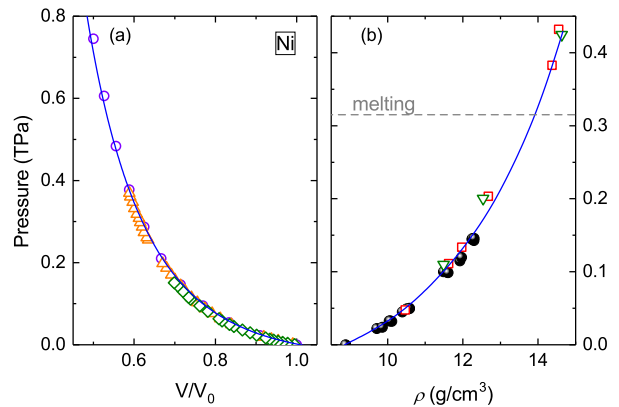}}
\caption{\label{fig4} The isotherm 300 K (a) and Hugoniot (b) of nickel from our calculation (solid lines) and experiment (diamonds \cite{A51}, triangles \cite{A12}, filled circles \cite{A53}, squares \cite{A54}; turned triangles \cite{A55}, open circles: EOS by Kormer et al \cite{A52}). The horizontal dashed line shows the approximate position of the melting point on the Hugoniot by data from our calculations (see text below).}
\end{figure}

\begin{figure}
\centering{
\includegraphics[width=8.0cm]{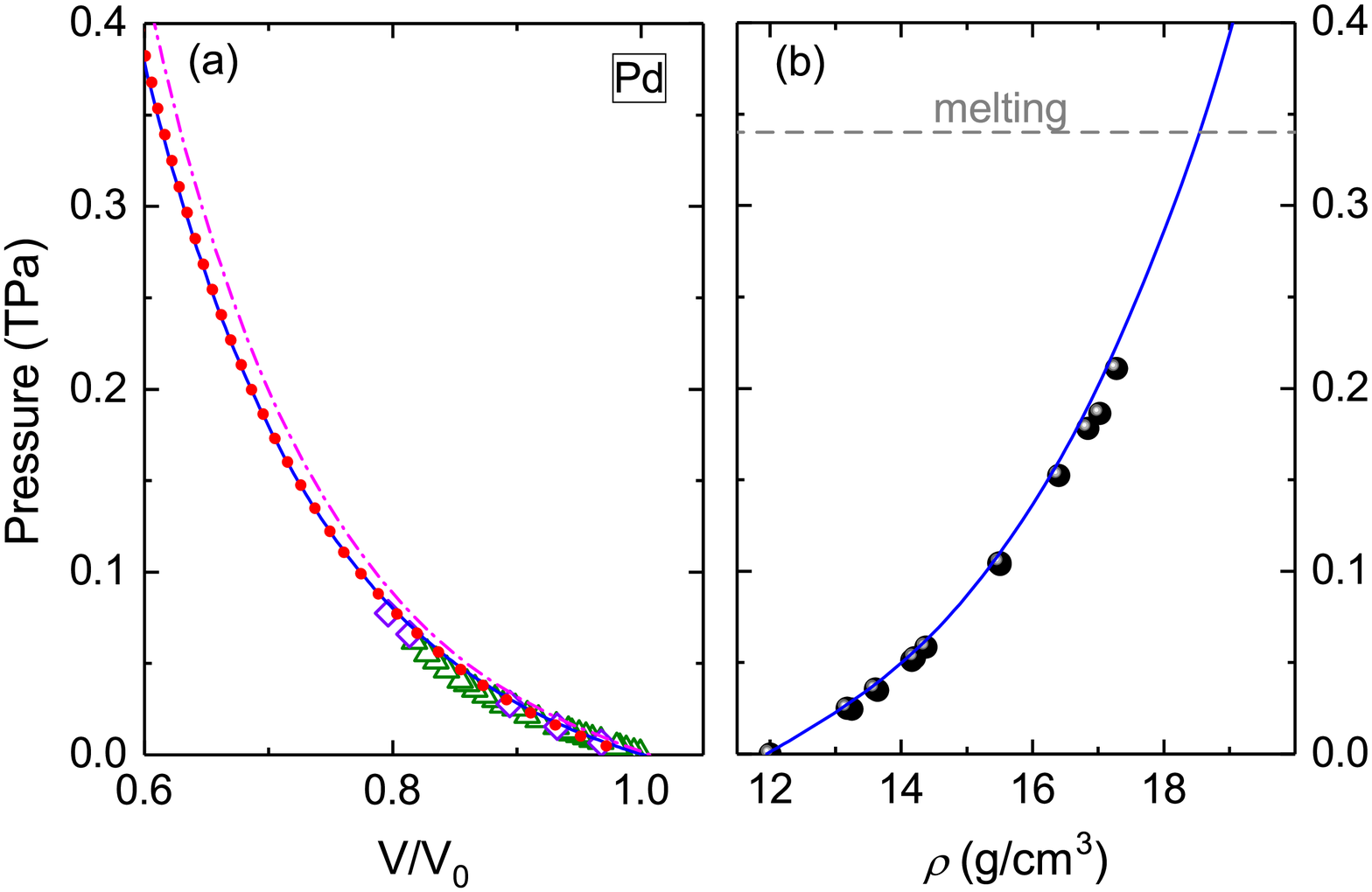}}
\caption{\label{fig5} The isotherm 300 K (a) and Hugoniot (b) of palladium from calculation (solid lines for this work, dash-dotted line \cite{A24}, red dots \cite{A23}) and experiment (diamonds \cite{A25}, triangles \cite{A26}, circles \cite{A53}). The horizontal dashed line shows the approximate position of the melting point on the Hugoniot by data from our calculations.}
\end{figure}

Consider the isothermal and shock compression of Ni, Pd, Rh, and Ir in comparison with available experimental data and other \emph{ab initio} calculations. Hereafter $V_0$ in all figures stands for the experimental specific volume at room temperature, i.e., $V_0$$=$73.82 a.u.$^3$ for Ni, $V_0$$=$99.37 a.u.$^3$ for Pd, $V_0$$=$92.83 a.u.$^3$ for Rh, and $V_0$$=$95.44 a.u.$^3$ for Ir \cite{A50}. Figure~\ref{fig4} shows results for nickel. The calculated isotherm 300 K is seen to agree well with available experimental data up to high pressures. The Hugoniot (Fig.~\ref{fig4}b) also agrees quite well with experiment. The same is true for palladium (Fig.~\ref{fig5}). It should however be noted that \emph{ab initio} calculations \cite{A24} give a somewhat overestimated value of the pressure on the isotherm for compressed Pd in comparison with our results \cite{A23}. Calculations presented here and in Ref. \cite{A23} were done in the GGA approximation for the XC functional, while those in paper \cite{A24} were performed with the LDA approximation which together with other factors could be of effect.

\begin{figure}
\centering{
\includegraphics[width=8.0cm]{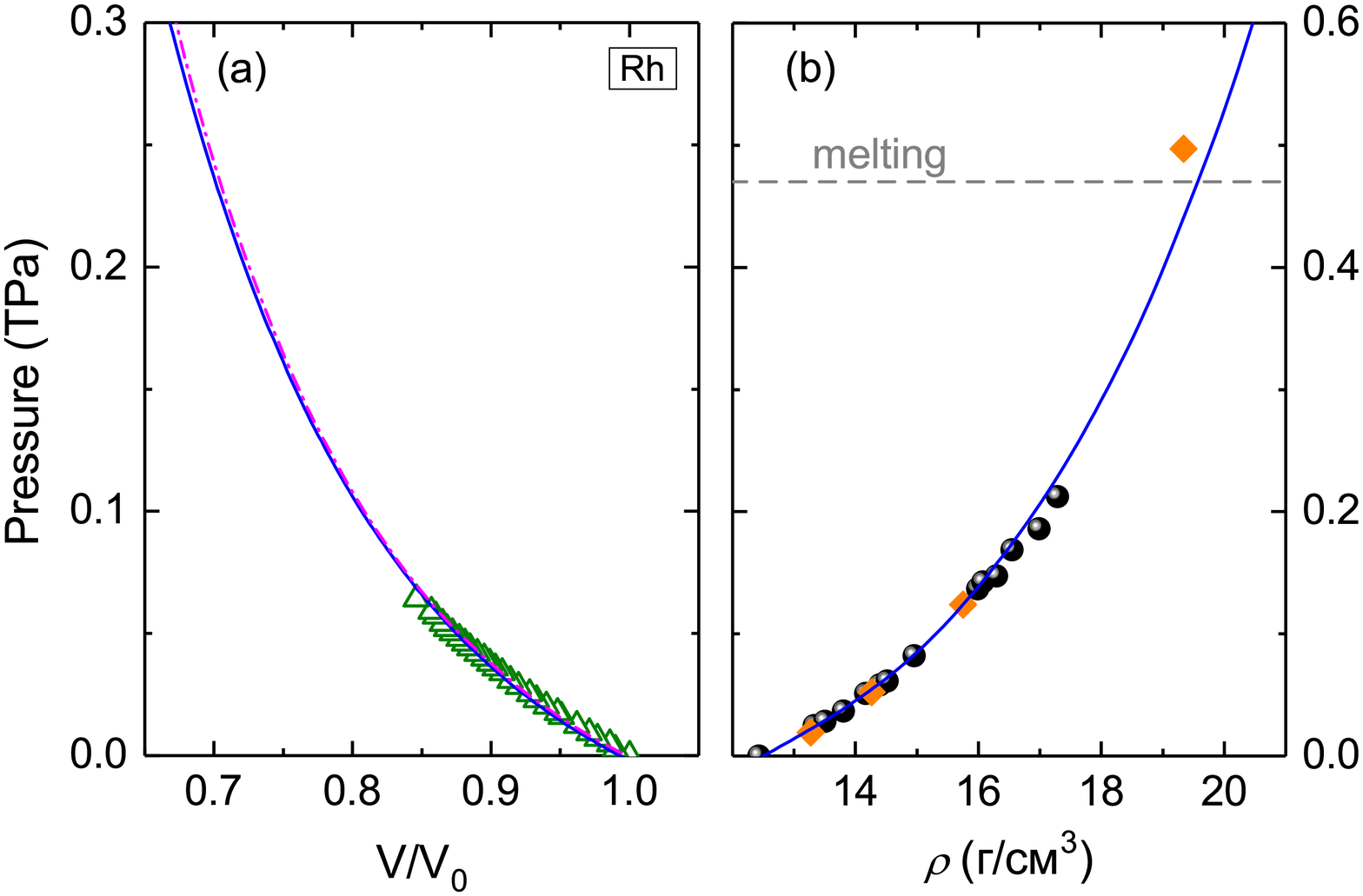}}
\caption{\label{fig6} The isotherm 300 K (a) and Hugoniot (b) of rhodium from calculation (solid lines for this work, dash-dotted line \cite{A23}) and experiment (triangles \cite{A27}, circles \cite{A53}, diamonds \cite{A54}). The horizontal dashed line shows the approximate position of the melting point on the Hugoniot by data from our calculations.}
\end{figure}

\begin{figure}
\centering{
\includegraphics[width=8.0cm]{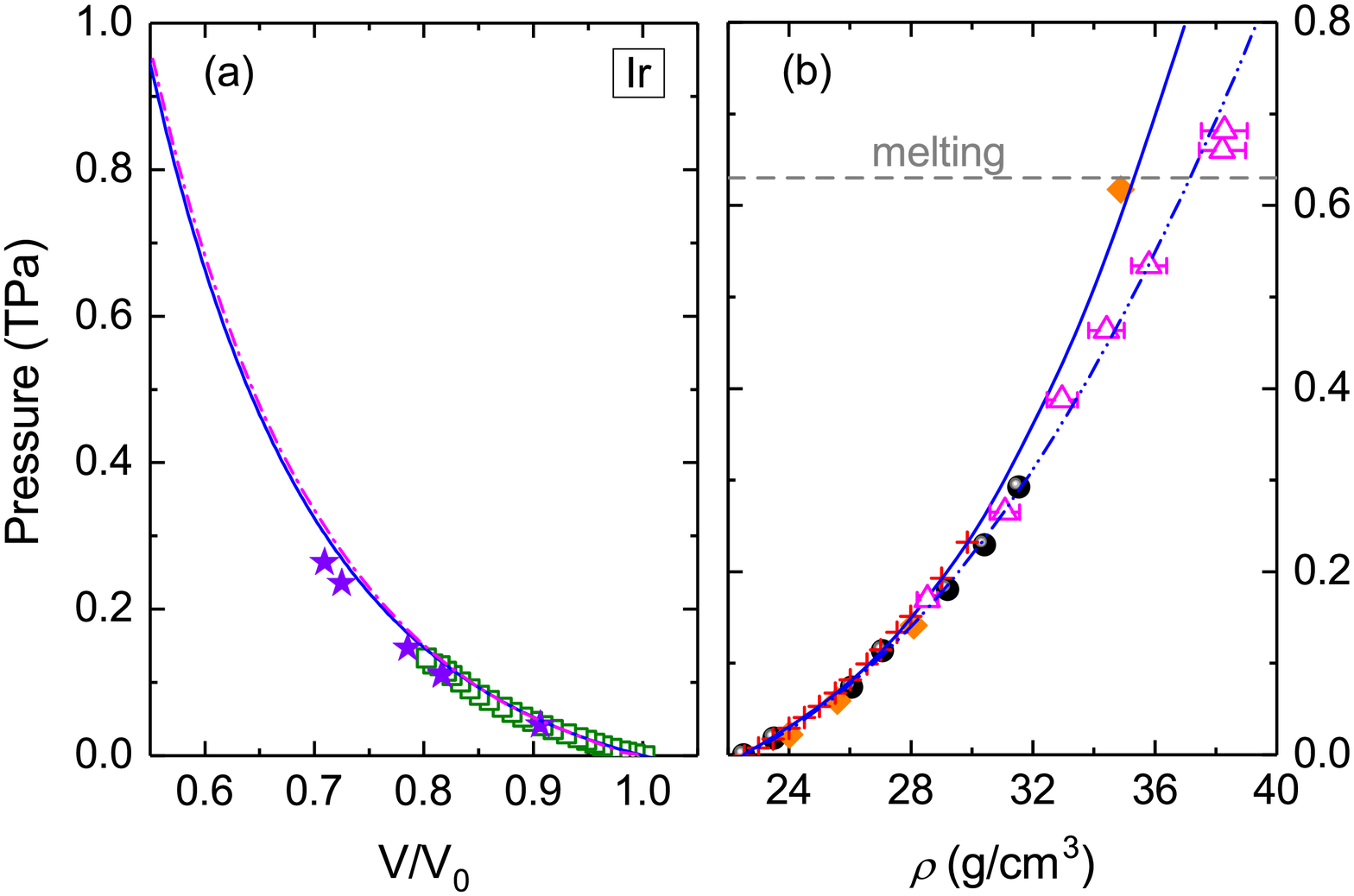}}
\caption{\label{fig7} The isotherm 300 K (a) and Hugoniot (b) of iridium from calculation (solid lines for this work, dash-dotted line \cite{A34}, pluses \cite{A35}, the dash-dot-dotted line on the right panel: our calculation of the isotherm 300 K) and experiment (pentagons \cite{A31}, squares \cite{A30}, diamonds \cite{A58}, circles \cite{A53}, triangles \cite{A57}). The horizontal dashed line shows the approximate position of the melting point on the Hugoniot by data from our calculations.}
\end{figure}

Figure~\ref{fig6} illustrates the isothermal and shock compression of rhodium. The calculated isotherm 300 K agrees well with experimental data from  \cite{A27}. Also seen is a good agreement with \emph{ab initio} calculations from  \cite{A23} despite that the authors used another gradient XC functional  \cite{A56}. The Hugoniot (Fig.~\ref{fig6}b) is also in quite a good agreement with experiment.

Figure~\ref{fig7} demonstrates the comparison for iridium. The isotherm agrees rather well with experiment. At pressures above 0.2 TPa, it is seen to somewhat deviate from experimental results presented in Ref. \cite{A31} and obtained with the double-stage diamond anvil technique, whose accuracy of measurement at high pressures remains under debate \cite{A59,A60}. For shock compression (Fig.~\ref{fig7}b), our results are seen to depart from experimental data \cite{A57} at $P$$>$0.3 TPa and to agree rather well with measurements \cite{A53,A58}. Our calculations are also in excellent agreement with \emph{ab initio} molecular dynamics data \cite{A35}. As for experimental results \cite{A57}, they raise some questions about the accuracy of measurements because if one draws the calculated isotherm 300 K of iridium on Fig.~\ref{fig7}b, it will be seen to pass, within the limits of error, through the experimental points \cite{A57}. The isotherms of the other phases (bcc, hcp, dhcp) are very close to that of the fcc structure (the differences are below 10 GPa) and it is unlikely that so large deviation of the Hugoniot can be caused by a structural transformation.

\begin{figure}
\centering{
\includegraphics[width=8.0cm]{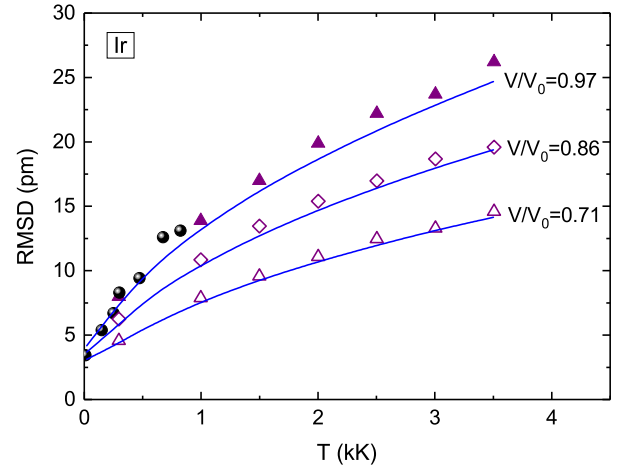}}
\caption{\label{fig8} Root-mean-square deviations of Ir atoms from fcc lattice equilibrium positions versus temperature for several specific volumes. Lines show our calculation, triangles and diamonds are QMD results \cite{A35}, and circles are data from constant pressure experiment \cite{A46}.}
\end{figure}

In paper \cite{A35}, QMD calculations were done to determine the root-mean-square deviations (RMSD) of Ir atoms from fcc lattice equilibrium positions at different specific volumes and temperatures. Figure~\ref{fig8} compares RMSD from \cite{A35} obtained with full anharmonism and the values calculated in this work in quasiharmonic approximation. It is seen that at $V/V_0$$=$0.97 and temperatures above 1 kK, our values lie a bit below the QMD results but the difference is no higher than 6\%. With the increasing compression, the difference becomes markedly smaller because under pressure the contribution of high-order anharmonicity effects gradually reduces. At $V/V_0$$=$0.71 ($P$$\gtrsim$0.3 TPa), our root-mean-square deviations are in very good agreement with QMD data.

\begin{figure}
\centering{
\includegraphics[width=8.0cm]{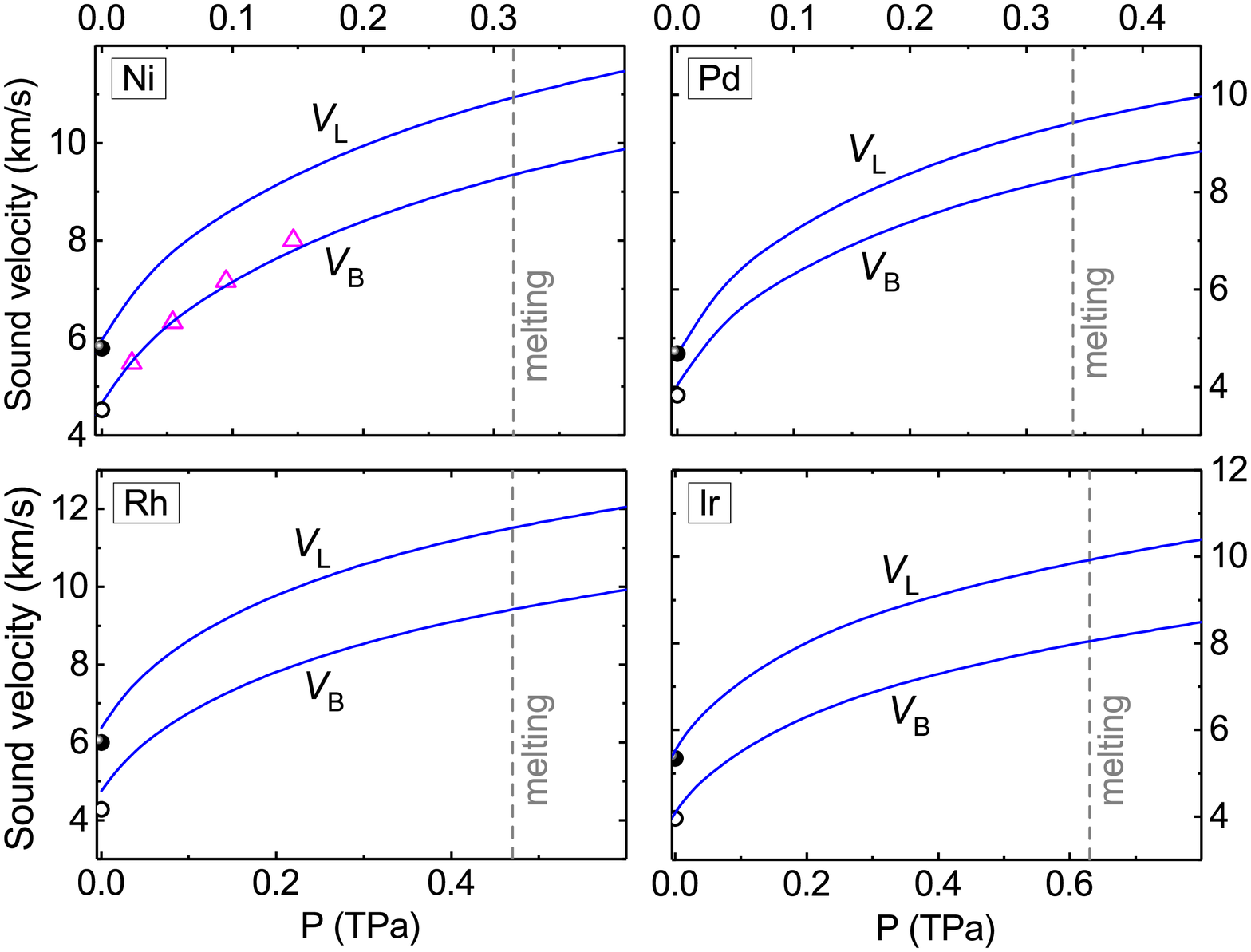}}
\caption{\label{fig9} Longitudinal ($V_L$) and bulk ($V_B$) sound velocities in shock compressed Ni, Pd, Rh, and Ir. The solid lines show calculations done in this work, the circles are for experimental data from \cite{A53}, and the triangles show data obtained from the experimental EOS at $T$$=$0 K \cite{A52}. The vertical dashed line is the approximate boundary of melting by data from our calculations.}
\end{figure}

\begin{figure}
\centering{
\includegraphics[width=8.0cm]{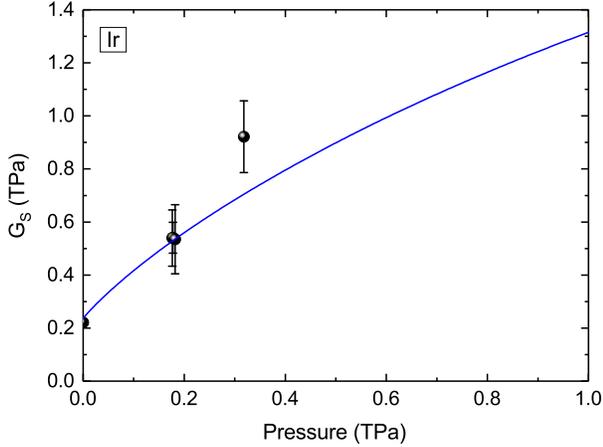}}
\caption{\label{fig10} Shear modulus of Ir versus pressure from our calculation (the solid line) and experiment (circles \cite{A61}).The value of $G_S$($P$$=$0) was taken from Ref. \cite{A62}.}
\end{figure}

The calculated elastic constants were used to determine sound velocities in the considered metals with the fcc structure under shock compression. The values of elastic constants at different compressions are provided in the Appendix. Our calculations of the elastic constants suggest that the fcc structure of all metals under consideration remains dynamically stable at least to 2 TPa (see the Appendix). Figure~\ref{fig9} shows the calculated longitudinal and bulk sound velocities of polycrystal Ni, Pd, Rh, and Ir versus pressure on the Hugoniot. Their experimental values are most often known only for ambient conditions where agreement with calculation is good (Fig.~\ref{fig9}). For nickel, data on bulk sound velocities, $V_B$, are available. They are obtained from an experimental EOS \cite{A52} at zero temperature and also agree well with our calculations. From the calculated elastic constants we can determine the shear modulus $G_S$ for polycrystal \cite{A48,A49}. Figure~\ref{fig10} compares $G_S$($P$) calculated in this work for iridium with measurements \cite{A61} taken under quasi-isentropic conditions. Up to $\sim$0.2 TPa, they agree very well. The experimental point at a higher pressure noticeably deviates from our function $G_S$($P$) which can be caused by the higher error of measurements \cite{A61} under high loading. Further measurements and calculations are needed to clear up the situation.

\begin{figure}
\centering{
\includegraphics[width=8.0cm]{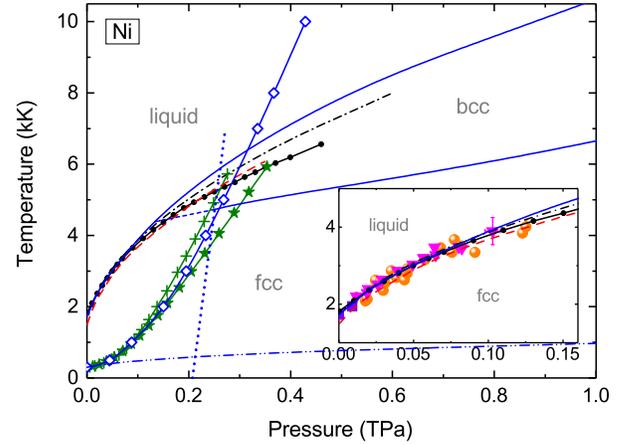}}
\caption{\label{fig11} The $PT$-diagram of nickel. The blue lines show our calculations: the solid ones are phase boundaries; the line with diamonds is the Hugoniot; the dash-dot-dotted line is the principal isentrope; and the dotted line is the dynamic stability boundary of the bcc phase. Melting curve data from other sources: the dashed-dotted line based on EOS \cite{A63}; the dashed line from classical MD calculations \cite{A64}; the line with black dots based on EOS \cite{A20}. Hugoniots: pluses \cite{A63} and pentagons \cite{A20}. The insert shows the melting curves with experimental points at relatively low $P$: squares \cite{A28}, circles \cite{A21}, and turned triangles \cite{A22}.}
\end{figure}

\begin{figure}
\centering{
\includegraphics[width=8.0cm]{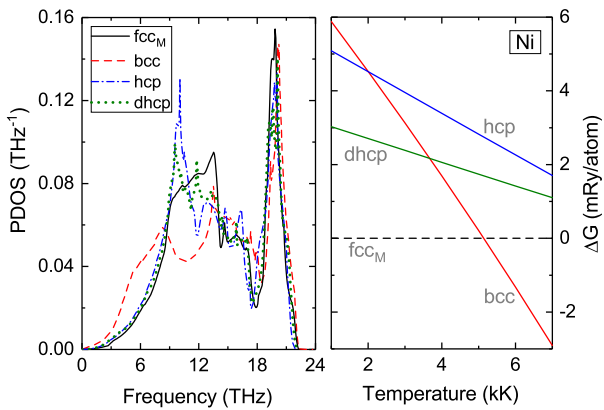}}
\caption{\label{fig12} Phonon density of states at $T$$=$0 K and $V/V_0$$=$0.6 (the left panel) and Gibbs energy difference (the right panel) for different structures of nickel at $P$$=$0.35 TPa.}
\end{figure}

Figure~\ref{fig11} presents the phase diagram of nickel obtained in this work, along with available calculated and experimental results by other authors. Our calculations show that at high pressures and temperatures the bcc structure of nickel is thermodynamically more stable than its fcc structure. The bcc phase stabilizes at $P$$>$0.14 and $T$$>$4 kK, which strongly resembles the recently discovered fcc-bcc transformation in copper and silver \cite{A05,A06,A08,A65}. Despite that at $T$$=$0 K the bcc structure has the energy higher than the close-packed phases have (see Fig.~\ref{fig2}), it has softer low-frequency phonon modes. Figure~\ref{fig12} (the left panel) shows the phonon densities of states (PDOS) at $V/V_0$$=$0.6 ($P$$\approx$0.35 TPa) and $T$$=$0 K for different structures of nickel. Its bcc phase is seen to have a higher PDOS in the low-frequency part of the phonon spectrum compared to the other phases. As a result, the Gibbs energy of the bcc phase with the increasing temperature becomes lower than that of the other phases which is seen in the right panel of Fig.~\ref{fig12}. The transition occurs due to higher entropy and a smaller contribution from lattice vibrations to energy compared to other lattices.

It is seen from Fig.~\ref{fig11} that the fcc-bcc boundary crosses the Hugoniot. That is, like in copper and silver, the fcc$\rightarrow$bcc transition in nickel can be detected in state-of-the-art shock experiments which can determine the crystal structure of investigated materials \cite{A04,A05,A06,A65}. By our estimates, the transition occur on the Hugoniot at $P$$\gtrsim$0.25 TPa. Unfortunately, the bcc structure at $T$$=$0 K loses its dynamic stability below certain pressure and at low compression it becomes impossible to determine its free energy within quasiharmonic approximation. In Figure~\ref{fig11}, the dotted line is the boundary of the dynamic stability of the bcc structure, and the fcc-bcc phase boundary line to the left of this curve is extrapolated to lower pressures (the short dashed line). To determine the phase boundary line at low $P$, more accurate calculations are required with consideration of all anharmonic effects.

The insert in Fig.~\ref{fig11} compares the melting curve $T_m$($P$) obtained in this work from the Lindemann criterion with available experimental data. The calculated $T_m$($P$) agrees quite well with experiment. The curves obtained by other authors either from MD calculations \cite{A64} or from semiempirical EOSs \cite{A20,A63} are also in quit a good agreement with our calculations but above 0.15 TPa they begin to depart from each other. However note that the structural transition to a new high pressure phase (bcc) must result in an increase of the melting point compared to a low pressure phase (fcc) \cite{A66}. Our calculations suggest that melting in shock-compressed nickel must occur at $P$$\thickapprox$0.315 TPa and $T$$\thickapprox$6.4 kK. Judging by the position of the fcc-bcc phase boundary, the structural transition will be difficult to observe in ramp experiments if the $P$-$T$ trajectory during measurements passes close enough to the principal isentrope (see Fig.~\ref{fig11}).

\begin{figure}
\centering{
\includegraphics[width=8.0cm]{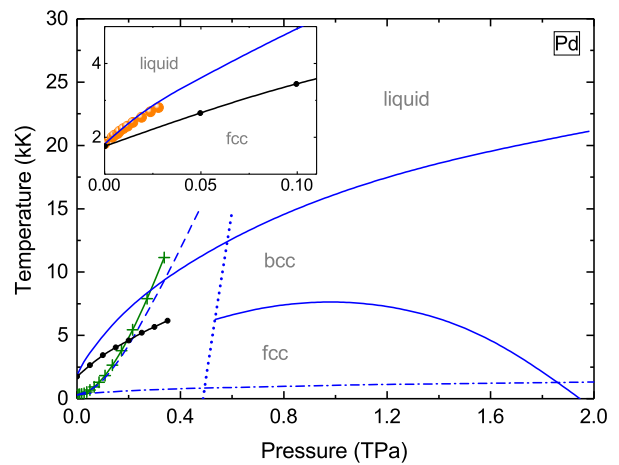}}
\caption{\label{fig13} The $PT$-diagram of palladium. Blue lines from our calculation: phase boundaries (solid lines), the Hugoniot (dashed line), the principal isentrope (dash-dotted line), and the bcc dynamic stability boundary (dotted line). The line with pluses shows the Hugoniot in quasiharmonic approximation \cite{A24}. The melting curves: from classical MD \cite{A24} (black dots connected by lines), and experimental data \cite{A17} (orange circles).}
\end{figure}

Now consider the phase diagram of palladium calculated in this work (Fig.~\ref{fig13}). It was stated earlier that unlike nickel, Pd has a fcc$\rightarrow$bcc structural transition at zero temperature and a pressure of about 1.9 TPa. As shown in Fig.~\ref{fig13}, with the increasing temperature the fcc-bcc phase boundary strongly shifts towards low pressures. As a result, at high temperatures the region of thermodynamic stability of the bcc phase extends well below a pressure of 1.9 TPa. The behavior of the fcc-bcc boundary is rather similar to that of platinum determined earlier in paper \cite{A08}. But unlike Pt, this line for palladium goes down in pressure much lower than 1.5 TPa. At $T$$=$0 K, bcc Pd becomes dynamically stable at much lower pressures compared to Pt which leads to the result mentioned above.

Like in nickel, there also exists a bcc dynamic stability boundary in palladium at relatively low pressures ($<$0.5 TPa) (the dotted line in Fig.~\ref{fig13}). To its left, bcc palladium is dynamically unstable at $T$$=$0 which does not allow the determination of the fcc-bcc boundary for low compressions in quasiharmonic approximation. More accurate calculations with full anharmonism are needed \cite{A66}. Accounting for the additional entropy contribution from high-order anharmonicity effects may change the potential energy surface and stabilize the bcc structure in this region of the phase diagram \cite{A67}. Nevertheless, the entire course of the phase boundary also allows us to suppose that under certain conditions the region where the bcc structure exists may get much lower than 0.5 TPa and intersect the Hugoniot (Fig.~\ref{fig13}). Like in nickel, the intersection is very likely to occur at pressures above 0.25 TPa.

Note that the melting curve obtained in this work agrees well with experiment \cite{A17} (see the insert in Fig.~\ref{fig13}). Calculations with classical molecular dynamics \cite{A24} give a curve that lies much lower than the experimental points and our curve. By our estimates, shocked Pd must melt at $P$$\approx$0.34 TPa and $T$$\approx$ 9.4 kK. With the existing estimations for temperatures from ramp compression experiments for different metals (2-4 kK for 0.2$<$$P$$<$1 TPa) \cite{A07,A68}, it seems to be quite difficult to detect a new bcc phase of palladium in similar experiments below 1 TPa.

\begin{figure}
\centering{
\includegraphics[width=8.0cm]{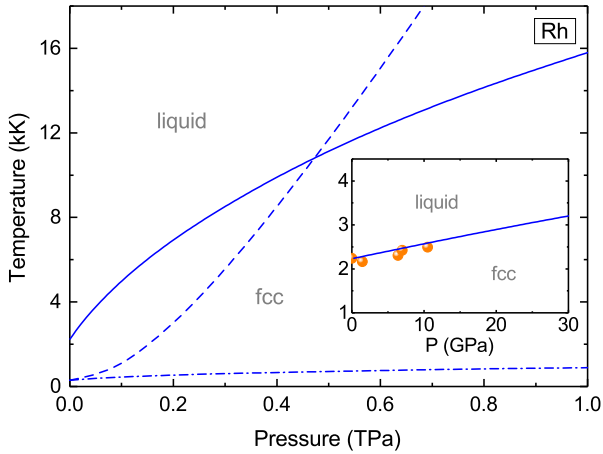}}
\caption{\label{fig14} The $PT$-diagram of rhodium. The solid line is calculated phase boundaries, dashed line is the Hugoniot, and dash-dotted line is the principal isentrope. Circles show the melting curve from experiment \cite{A28}.}
\end{figure}

\begin{figure}
\centering{
\includegraphics[width=8.0cm]{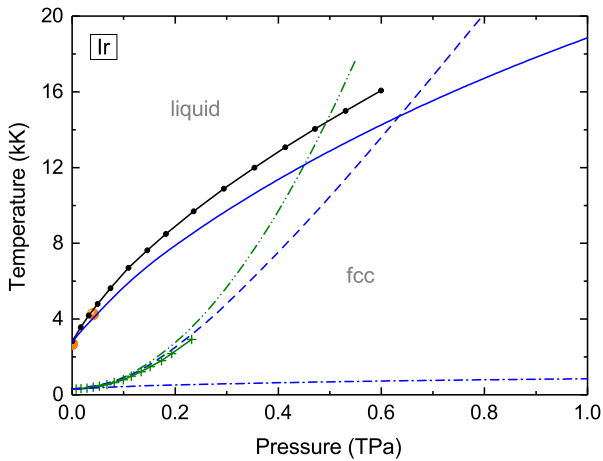}}
\caption{\label{fig15} The $PT$-diagram of iridium. Blue lines show our calculations: phase boundaries (solid lines), the Hugoniot (dashed line), and the principal isentrope (dash-dotted line). The black dots connected by lines and the dash-dot-dotted line are the melting curve and Hugoniot from QMD calculations \cite{A34}. The line with crosses is the Hugoniot from QMD calculations \cite{A35}. The orange circles are experimental points on the melting curve \cite{A32}.}
\end{figure}

\begin{figure}
\centering{
\includegraphics[width=8.0cm]{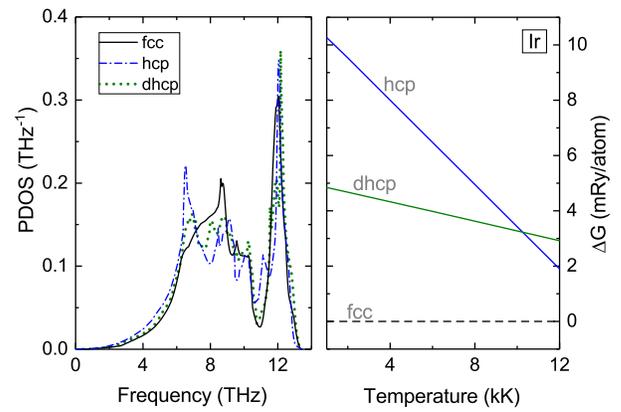}}
\caption{\label{fig16} Phonon densities of states at $T$$=$0 K and $V/V_0$$=$0.7 (the left panel) and Gibbs energy differences (the right panel) for different iridium structures at $P$$=$0.32 TPa.}
\end{figure}

Figures \ref{fig14} and \ref{fig15} present phase diagrams calculated in this work for rhodium and iridium. The calculated melting curve of rhodium (the insert in Fig.~\ref{fig14}) agrees well with available experimental data. Our study did not reveal on the $PT$-diagram of Rh a region where could exist a phase other than fcc below 1 TPa (fig.~\ref{fig14}). At zero temperature, bcc Rh remains dynamically unstable at least to pressures about 3 TPa. Our calculations suggest that rhodium crystal under shock compression begins to melt at $P$$\approx$0.47 TPa and $T$$\approx$10.8 kK.

We also did not find any structural transitions in iridium on its $PT$-diagram (Fig.~\ref{fig15}). Like in rhodium, its bcc structure is dynamically unstable at $T$$=$0 K to pressures of a few TPa. Calculations done in this work suggest that the hexagonal close-packed structures are not energetically more preferable than the fcc phase at high $P$ and $T$. Figure~\ref{fig16} presents the phonon DOS and Gibbs energy differences calculated for iridium phases at $P$$=$0.32 TPa. PDOS for the close-packed structures are seen to run quite close to each other. As a result, with the increasing temperature the Gibbs energy of the fcc phase remains the lowest up to melting (the left panel of Fig.~\ref{fig16}). According to \cite{A34}, the energy of the dhcp structure is lower than that of the fcc one at high pressures and temperatures. However, unlike \cite{A34}, we have not succeeded to observe a transition to the dhcp structure in iridium with the increasing temperature of crystal under high pressures. Note that the same authors earlier predicted a structural transition from fcc to rhcp to occur in platinum \cite{A69}. However the transition was not observed either in dynamic or static experiments \cite{A05,A70}.

The melting curve calculated in this work agrees well with experiment \cite{A32}. QMD calculations \cite{A34} by the Z method give higher melting temperatures with the increasing pressure. Our calculations suggest that the melting of shocked iridium must occur at $P$$\approx$0.63 TPa and $T$$\approx$14.7 kK. Our Hugoniot is in good agreement with QMD results \cite{A35} (Fig.~\ref{fig15}). Calculations from \cite{A34} give a steeper dependence of temperature on pressure under shock compression and, accordingly, a lower value of pressure (by about 30\%) at which iridium begins to melt. It is clear that further MD calculations are needed to describe more accurately the behavior of the melting curve and Hugoniot of iridium at high $P$ and $T$.

\section{Conclusion}

The structural stability and thermodynamic and elastic properties of Ni, Pd, Rh, and Ir were studied in this work with the all-electron full-potential LMTO method. From the calculated elastic constants, it is shown for all the metals that their fcc structure remains dynamically stable at least to pressures about 2 TPa. Calculations were done to determine the isotherms, Hugoniots in $P$-$\rho$ and $P$-$T$ coordinates, and sound velocities on the Hugoniot. Melting curves were calculated from the Lindemann criterion and parameters of melting under shock compression were estimated to be $P$$\approx$0.315 TPa and $T$$\approx$6.4 kK for Ni, $P$$\approx$0.34 TPa and $T$$\approx$9.4 kK for Pd, $P$$\approx$0.47 TPa and $T$$\approx$10.8 kK for Rh, and $P$$\approx$0.63 TPa and $T$$\approx$14.7 kK for Ir. Calculated results are shown to agree well with currently available experimental data.

It is shown by calculations for nickel and palladium that their bcc phase be most thermodynamically stable under high pressures and temperatures instead of the fcc phase that is stable near the ambient conditions. It was determined for nickel that its Hugoniot intersect the fcc-bcc phase boundary at a pressure of about 0.25 TPa. Results for palladium are not so unambiguous because of the shortcomings present in the quasiharmonic approximation we used. Here we can only hypothesize that the fcc$\rightarrow$bcc transition may also occur in palladium under shock compression above a couple of hundred GPa. As shown in our calculations, rhodium and iridium do not undergo any structural changes under pressures to a few TPa and temperatures up to melting. Their bcc structure remains dynamically unstable at least to 3 TPa.

The results we obtained in this work and in our earlier studies \cite{A03,A08} allow us to speak about a common tendency for the metals Ni, Pd, Pt, Cu, Ag, and Au. They all undergo a structural transition from their face-centered to body-centered cubic phase under high temperatures and pressures less than 2 TPa. For Cu, Ag, and Au, this transition has been recently observed in experiments \cite{A04,A05,A06,A07,A65}. Our calculations show that Pd, Pt, and Au at $P$$\leq$2 TPa undergo transition to their bcc phase at zero temperature as well. We predict that nickel under Earth-core conditions ($P$$\sim$0.3 TPa and $T$$\sim$6 kK) will have the bcc structure rather than fcc, as was expected. Possibly, the detection of the fcc$\rightarrow$bcc transition in nickel at high $P$ and $T$ will help more accurately judge about the structural state of the iron-nickel alloy in the Earth core interior.

\section{Appendix}

Tables~\ref{table1}-~\ref{table4} present calculated values of elastic constants at different compressions for Ni, Pd, Rh, and Ir.

\begin{table}
\caption{\label{table1}The elastic constants $C_{ij}$ and cold pressure $P_c$ (all in GPa) of nickel at different compressions. Experimental results under ambient conditions are marked as bold.}
\begin{ruledtabular}
\begin{tabular}{lllll}
$V/V_0$&$C_{11}$&$C_{12}$&$C_{44}$&$P_c$\\
\hline
1.05&222.0&124.2&110.5&-9.018 \\
1.0&274.0&162.4&132.5&-0.354 \\
\textbf{Exp. [71]}&\textbf{261.2}&\textbf{150.8}&\textbf{131.7}&\textbf{0.0} \\
0.9&420.3&272.4&192.6&26.72 \\
0.8&653.0&455.9&285.0&75.61 \\
0.7&1036.3&775.8&432.4&166.6 \\
0.6&1703.2&1361.7&675.7&343.4 \\
0.5&2932.6&2498.5&1025.1&709.8 \\
0.4&5550.1&4849.9&1823.4&1537.1 \\
0.35&8040.2&7156.9&2503.5&2365.1 \\
\end{tabular}
\end{ruledtabular}
\end{table}

\begin{table}
\caption{\label{table2}The elastic constants $C_{ij}$ and cold pressure $P_c$ (all in GPa) of palladium at different compressions. Experimental results under ambient conditions are marked as bold.}
\begin{ruledtabular}
\begin{tabular}{lllll}
$V/V_0$&$C_{11}$&$C_{12}$&$C_{44}$&$P_c$\\
\hline
1.05&178.7&134.6&56.5&-10.28 \\
1.0&231.7&177.5&73.6&-1.534 \\
\textbf{Exp. [72]}&\textbf{234.1}&\textbf{176.1}&\textbf{71.2}&\textbf{0.0} \\
0.9&388.8&305.7&123.2&26.35 \\
0.8&646.8&520.4&205.5&78.09 \\
0.7&1088.6&893.7&343.8&177.8 \\
0.6&1865.7&1570.9&584.7&376.6 \\
0.5&3370.7&2951.8&1025.4&798.6 \\
0.4&6484.3&5898.1&1916.6&1797.9 \\
\end{tabular}
\end{ruledtabular}
\end{table}

\begin{table}
\caption{\label{table3}The elastic constants $C_{ij}$ and cold pressure $P_c$ (all in GPa) of rhodium at different compressions.}
\begin{ruledtabular}
\begin{tabular}{lllll}
$V/V_0$&$C_{11}$&$C_{12}$&$C_{44}$&$P_c$\\
\hline
1.05&352.8&152.9&178.2&-15.81 \\
1.0&435.4&203.5&216.1&-3.653 \\
0.9&665.9&351.1&318.1&34.60 \\
0.8&1033.9&599.9&473.0&104.2 \\
0.7&1640.4&1037.9&713.2&234.8 \\
0.6&2675.9&1831.7&1088.6&488.8 \\
0.5&4543.6&3356.7&1734.1&1009.4 \\
0.4&8257.2&6641.8&2908.4&2194.8 \\
\end{tabular}
\end{ruledtabular}
\end{table}

\begin{table}
\caption{\label{table4}The elastic constants $C_{ij}$ and cold pressure $P_c$ (all in GPa) of iridium at different compressions. Experimental results under ambient conditions are marked as bold.}
\begin{ruledtabular}
\begin{tabular}{lllll}
$V/V_0$&$C_{11}$&$C_{12}$&$C_{44}$&$P_c$\\
\hline
1.05&498.5&194.1&232.4&-17.59 \\
1.0&614.0&262.5&284.0&-1.184 \\
\textbf{Exp. [73]}&\textbf{596}&\textbf{252}&\textbf{270}&\textbf{0.0} \\
0.9&938.0&465.7&424.9&50.82 \\
0.8&1442.9&801.4&640.6&145.9 \\
0.7&2255.7&1377.1&977.9&322.3 \\
0.6&3620.3&2412.0&1517.5&661.6 \\
0.5&5989.9&4373.0&2369.8&1346.8 \\
0.4&10849.1&8701.3&3837.7&2899.4 \\
\end{tabular}
\end{ruledtabular}
\end{table}

\end{document}